# Spin torque switching with the giant spin Hall effect of tantalum

Luqiao Liu[*,1], Chi-Feng Pai[*,1], Y. Li[1], H. W. Tseng[1], D. C. Ralph[1,2] and R. A. Buhrman[1]

[1]Cornell University and [2]Kavli Institute at Cornell, Ithaca, New York, 14853

We report a giant spin Hall effect (SHE) in $\beta$-Ta that generates spin currents intense enough to induce efficient spin-transfer-torque switching of ferromagnets, thereby providing a new approach for controlling magnetic devices that can be superior to existing technologies. We quantify this SHE by three independent methods and demonstrate spin-torque (ST) switching of both out-of-plane and in-plane magnetized layers. We implement a three-terminal device that utilizes current passing through a low impedance Ta-ferromagnet bilayer to effect switching of a nanomagnet, with a higher-impedance magnetic tunnel junction for read-out. The efficiency and reliability of this device, together with its simplicity of fabrication, suggest that this three-terminal SHE-ST design can eliminate the main obstacles currently impeding the development of magnetic memory and non-volatile spin logic technologies.

---

[*] These authors contributed equally to this work.

Spin-polarized currents can be used to apply torques to magnetic moments by direct transfer of spin angular momentum, thereby enabling manipulation of nanoscale magnetic devices using currents that are orders of magnitude lower than required for magnetic-field-based control (*1-5*). So far the only way to generate spin currents strong enough for spin-torque manipulation of magnets in practical applications has been to send an electron current through a magnetic polarizing layer, with the result that the most promising device geometry for applications has been the two-terminal magnetic tunnel junction (MTJ), possessing a ferromagnetic layer (FM)/tunnel barrier/FM structure. MTJs can have excellent spin torque efficiency, but are proving challenging to operate reliably; it is difficult to manufacture large-scale memories in which enough spin current can pass through the tunnel barriers to drive reliable magnetic switching without occasionally damaging a barrier. Achieving reliable reading of the MTJ resistance without ever causing switching during a read step is also a challenge. It has been known for some time that a spin current can, alternatively, be generated in non-magnetic materials by the spin Hall effect (SHE) (*6-12*), in which spin-orbit coupling causes electrons with different spins to deflect in different directions yielding a pure spin current transverse to an applied charge current. However, very few attempts have been carried out to utilize this spin current for manipulating magnetic moments (*13, 14*). Here we report the discovery of a giant SHE in the high resistivity form of tantalum ($\beta$–Ta) (*15*), and we demonstrate that this allows an electrical current in a thin Ta layer to efficiently induce spin-torque switching of an adjacent thin film ferromagnet, for both perpendicular-to-plane and in-plane magnetized samples at room temperature. We have quantified the magnitude of the SHE in Ta using three different methods and find the spin Hall angle to be $\theta_{SH}^{Ta}$ = 0.12-0.15, larger than and with a sign opposite to the spin Hall angle of Pt, $\left|\theta_{SH}^{Pt}\right| \approx 0.07$ (*16-21*), and comparable to one report for Pt-doped Au, $\left|\theta_{SH}^{Au(Pt)}\right| = 0.12 \pm 0.04$ (22). (Here $\theta_{SH} = J_S / J_e$, where $J_e$ is the charge current density and $[\hbar/(2e)]J_S$ is the spin current density arising from the SHE.) Unlike Pt, Ta does not significantly increase the magnetic damping (energy dissipation) in an adjacent thin-film magnet, which has the consequence that the giant spin Hall effect spin torque (SHE-ST) from Ta can be uniquely effective in driving magnetic reversal of in-plane-polarized magnetic layers via an "anti-damping" spin-torque mechanism (5). We employ this effect to implement a novel three-terminal device geometry in which the SHE-ST from Ta produces current-induced switching of in-plane polarized CoFeB layer, with read-out using a magnetic tunnel junction with a large magnetoresistance. This geometry is straightforward to fabricate and can have comparable efficiency to conventional two-terminal MTJs while providing greatly improved reliability and output signal levels, and therefore offers a superior approach for magnetic memory and non-volatile spin logic applications.

High resistivity $\beta$–Ta is produced when Ta is sputter deposited or evaporated onto amorphous surfaces such as, *e.g.,* oxidized Si (23) or CoFeB. Tanaka *et al.* (24) have predicted, based on an ab initio calculation, that highly resistive Ta may have a large spin Hall angle, comparable to or greater than that of Pt, and with the opposite sign in comparison to Pt or Au. In contrast, an experiment by Morota *et al.* that utilized a non-local spin valve measurement (25) reported a very low value for the Ta spin Hall angle, 0.0037. However, as recently explained (20, 21), this measurement technique can produce a large underestimate of the spin Hall angle because the analysis in (25) does not correctly account for how charge flow in the spin Hall material is shunted by a highly conducting Cu electrode at the cross point of the non-local spin valve bridge. This problem is most severe for spin Hall materials that are highly resistive, as is Ta.

**Giant spin Hall effect in Ta.** We measured the SHE in $\beta$–Ta using the ST induced ferromagnetic resonance (ST-FMR) technique, previously introduced in studies of Pt (*20*). Our samples consisted of Co$_{40}$Fe$_{40}$B$_{20}$(4)/Ta(8) (thickness in nm) bilayers sputter-deposited onto oxidized Si substrates and patterned into 10 μm wide strips. Measurements of the bilayer resistance as a function of varying Ta thickness determined that the Ta resistivity was $\rho_{Ta} \approx 190$ μΩ-cm, confirming the $\beta$–Ta phase. The Co$_{40}$Fe$_{40}$B$_{20}$ resistivity was $\rho_{CoFeB} \approx 170$ μΩ-cm and for a 4-nm thick film the magnetic moment was oriented in-plane. We applied an oscillating current $I_{RF}$ along the strips in the current-in-plane configuration, with an external magnetic field $B_{ext}$ in the film plane at a 45° angle with respect to the current direction (Fig. 1A). Because of the SHE, the oscillating current in the Ta generated an oscillating spin current that flowed perpendicular to the sample plane and exerted an oscillating spin torque on the magnetic moment of the CoFeB layer. When the frequency of the bias current and the magnitude of the bias magnetic field satisfied the ferromagnetic resonance condition, magnetic precession occurred. This resulted in a measurable DC voltage due to the mixing of $I_{RF}$ and the oscillating anisotropic magnetoresistance (AMR) of the CoFeB. A typical resonance signal is shown in Fig. 1B, where the resonance peak is fitted by the sum of a symmetric Lorentzian and an antisymmetric Lorentzian. The symmetric component of the peak arises from the SHE-ST while the antisymmetric peak is due to the torque generated by the Oersted field from the current in the Ta, with the difference in lineshape being due to the two torques' orthogonal directions (Fig. 1A) [for a detailed discussion of the lineshapes see (*20, 21*)]. We measured the resonant signal for different bias frequencies and determined that the positions of the resonant peaks agree well with the Kittel formula $f = (\gamma/2\pi)[B(B + \mu_0 M_{eff})]^{1/2}$ (see Fig. 1B inset), where $\gamma = 1.76 \times 10^{11}$ HzT$^{-1}$ is the gyromagnetic ratio and $\mu_0 M_{eff} = 1.3$ T is the effective demagnetization field.

To compare the SHE in Ta with that of Pt, we made and measured a different sample with the stack structure: substrate/CoFeB(3)/Pt(6) (thicknesses in nm), with the result shown in Fig. 1C. Comparing the resonant signals of CoFeB/Ta and CoFeB/Pt in Fig. 1B and C, we see that the antisymmetric peaks of the two samples have the same sign, as expected from their common origin being the Oersted field from the current flowing in the non-magnetic (NM) layer (*26*). The symmetric peaks in the two cases are opposite in sign, which directly shows that the SHE in Ta is opposite to that in Pt, in agreement with the prediction (*24*) and the previous measurement (*25*).

We measured the magnitude of the SHE using a self-calibrated technique that uses the ratio of the symmetric peak amplitude *S* to the antisymmetric peak *A* to determine the strength of the spin Hall torque relative to the Oersted-field torque [see Ref. (*20*)]. Independent of the frequency employed, we found the consistent result that $J_S/J_e = 0.15 \pm 0.03$ in our 8 nm Ta films, a value larger than for any other material reported previously. This value of $J_S/J_e$ represents the spin Hall angle $\theta_{SH}$ if the spin diffusion length $\lambda_{sf}$ in Ta is much less than the Ta thickness (*20, 21*). If $\lambda_{sf}$ is comparable to or larger than the film thickness, then the bulk value of $\theta_{SH}$ is even larger than $0.15 \pm 0.03$.

If the spin torque from the SHE is to be utilized for switching nanomagnets by the conventional anti-damping ST switching mechanism (*5*), it is important that the NM layer does not substantially increase the effective magnetic damping of the adjacent FM by the spin pumping effect (*27, 28*). The ST-FMR measurements discussed above allow a determination of the Gilbert damping coefficient $\alpha$ from the linewidth $\Delta B$ (half width at half maximum) of the FMR peak, using the relationship $\alpha = (\gamma/2\pi f)\Delta B$. The results shown in Fig. 1D indicate that $\alpha = 0.008$ for CoFeB/Ta bilayer film, a number very close to the intrinsic value expected for a 4 nm thick CoFeB layer (*29*), while in comparison $\alpha$ is much larger, $\approx 0.025$, for the CoFeB/Pt sample. This is consistent with prior work (*27*) in which damping due to spin pumping was determined to be much stronger in FM/Pt bilayers than in FM/Ta, although the phase of Ta studied in ref (*27*) was not reported. Our observation of a strong spin Hall effect in $\beta$–Ta is not in conflict with the weakness of the spin pumping effect in Ta films, because the strength of the spin pumping depends not just on the strength of spin-orbit coupling, but also on the ratio of the elastic scattering time to the spin flip scattering time and the value of the spin mixing conductance (*28*), either or both of which might be smaller in $\beta$–Ta than Pt.

**Switching a perpendicularly magnetized ferromagnetic layer with the spin Hall effect.** Previous experiments (*14*) utilizing a perpendicularly magnetized FM deposited on Pt, and with a small magnetic field applied in the direction of the electrical current, have demonstrated that the SHE-ST will, once it is strong enough relative to the magnetic anisotropy field, abruptly rotate the out-of-plane moment from the nearly vertical positive (upwards) orientation to the nearly vertical negative (downwards) orientation, or *vice versa*, depending on the direction of the current flow and the SHE sign. We have verified that the stronger SHE in Ta can achieve the same switching effect but with the opposite sign compared to Pt. For this measurement we deposited a thin film stack with the structure: substrate/Ta(4)/CoFeB(1)/MgO(1.6)/Ta(1) (thicknesses in nm) and patterned it into Hall bars 2.5-20 μm wide and 3-200 μm long, as illustrated in the inset of Fig. 2A. MgO was employed as a capping layer because previous studies (*30*) have shown for sufficiently thin CoFeB that the Ta/CoFeB/MgO structure has a strong perpendicular magnetic anisotropy, as confirmed by our measurements (Fig. 2A). (The top Ta layer served merely to protect the MgO from degradation due to exposure to atmosphere.) For the ST switching measurement, a DC current was applied along the strip and the anomalous Hall resistance $R_H$ was recorded to monitor the change in the z component of the CoFeB magnetization since $R_H \propto M_z = M_S \sin\theta$. A static magnetic field $B_{ext}$ was applied almost parallel (or antiparallel) to the in-plane current direction while the angle $\beta$ between $B_{ext}$ and the film plane was kept fixed, initially at $\beta = 0°$. Figure 2B shows an example of the abrupt current-induced switching that occurs due to the SHE-ST, as measured for a 2.5 μm wide sample with $\beta = 0°$ and $B_{ext} = \pm 10$ mT. The switching curves shown are obtained under the same bias conditions as employed for Fig. 1 of Ref. (*14*), and inspection of that figure reveals that the switching direction caused by the in-plane current in Fig. 2B is opposite to that reported for the Pt/Co/AlO$_x$ system. To further verify the origin of the reversal of switching direction in those two samples we made additional control samples from a Pt/CoFeB/MgO multilayer and found that the switching direction is the same as with Pt/Co/AlO$_x$, demonstrating that the sign reversal comes from the difference between the sign of the SHE in Pt and Ta, and is not due to any differences between the FM/oxide interfaces or between Co and CoFeB.

To quantitatively determine the magnitude of the spin Hall angle from the response of perpendicularly magnetized Ta/CoFeB/MgO samples, a small field angle $\beta \approx 2°$ was employed while we swept the magnetic field. With a non-zero $\beta$, the z component of the external magnetic field $B_z = B_{ext} \sin\beta$ causes the magnetization of the Hall bar structure to remain uniformly magnetized as long as the current is well below the switching point, so that the magnetization rotates coherently with field and current as shown in Fig. 2C. For convenience in the data analysis, in the following we will treat $B_{ext}$ as function of $R_H$ instead of the reverse. As demonstrated in Ref. (*14*), the difference between the $B_{ext}(R_H)$ curves for $I = +0.7$ mA and $-0.7$ mA can be shown, within a macrospin model, to be proportional to the applied spin torque:
$\Delta B[R_H(\theta)] = B_+(\theta) - B_-(\theta) = 2\tau_{ST}^0 / \sin(\theta - \beta)$. Here $B_{+/-}(\theta)$ is defined as the value of $B_{ext}$ required to produce a given value of the magnetization angle $\theta$ when $I$ is positive/negative. Figure 2D shows $\Delta B(R_H)$ determined by subtracting the two data sets in Fig. 2C. We plot $R_H$ normalized with respect to its maximum value, so that it is equal to $\sin\theta$. Using a one-parameter fit, the magnitude of the spin torque can be determined to be $\tau_{ST}^0 \approx 2.1$ mT for $|I| = 0.7$ mA. The $\tau_{ST}^0 / I$ ratios obtained for different values of applied current are summarized in the inset of Fig. 2D, and on average we find $\tau_{ST}^0 / I \approx 2.8 \pm 0.6$ mT/mA. By using the formula $J_S = 2eM_S t\tau_{ST}^0 / \hbar$ with the parameters $M_S = (1.1 \pm 0.2) \times 10^6$ A/m and $t = 1.0 \pm 0.1$ nm, we obtain $J_S / J_e = 0.12 \pm 0.03$ for the 4 nm Ta layer, in quite reasonable accord with the value $J_S / J_e = 0.15 \pm 0.03$ from the ST-FMR study. Here we assume a uniform current density throughout both the Ta and CoFeB layers since their resistivities are similar, $\rho_{Ta} \approx 190$ μΩ-cm and $\rho_{CoFeB} \approx 170$ μΩ-cm [See Section S1 of the supporting online materials (SOM)].

**Spin torque switching of an in-plane polarized magnet using a three-terminal spin Hall device**. The giant SHE in Ta, together with its negligible effect on the damping of adjacent magnetic layers, makes Ta an excellent material for effecting ST switching of an in-plane magnetized nanomagnet. In conventional anti-damping ST switching where the spins are injected either nearly parallel or anti-parallel to the initial orientation of the local magnetic moment, the critical current density for switching in the absence of thermal fluctuations is given by (*3, 31*)

$$J_{C0} \approx \frac{2e}{\hbar} \mu_0 M_S t \alpha (H_C + M_{eff} / 2) / (J_S / J_e), \quad (1)$$

where $M_S$, $t$, and $H_C$ represent the saturation magnetization, the thickness and the coercive field of the FM nanomagnet, respectively.

To demonstrate in-plane magnetic switching induced by the SHE, we fabricated a three-terminal device, consisting of the multilayer: substrate/Ta(6.2)/CoFeB(1.6)/MgO(1.6)/

CoFeB(3.8)/Ta(5)/Ru(5) (thicknesses in nm) patterned into the geometry shown in Fig. 3A [see Methods section SOM]. The Ta bottom layer was patterned into a 1 μm wide and 5 μm long strip (with resistance 3 kΩ) and the rest of the layers were etched to form a magnetic tunnel junction (MTJ) nanopillar on top of the Ta with lateral dimensions ~ 100 × 350 nm, and with the long axis of the nanopillar perpendicular to the long axis of the Ta microstrip.

The magnetoresistance response of one of these MTJs is shown in Fig. 3B, which indicates a coercive field $B_C \approx 4$ mT, a zero bias MTJ resistance $R_{MTJ} \approx 65$ kΩ, and a tunneling magnetoresistance (TMR) ≈ 50%. During subsequent magnetic switching measurements we applied a -3.5 mT in-plane magnetic field along the long axis of the MTJ to cancel the dipole field from the top layer of the MTJ acting on the bottom layer, and thus biased the junction at the midpoint of its minor magnetoresistance loop. We then applied a DC current $I_{Ta}$ to the Ta microstrip while monitoring the differential resistance $dV/dI$ of the MTJ (Fig. 3A). Figure 3C shows that abrupt hysteretic switching of the MTJ resistance occurred when $I_{Ta}$ was swept through 1 mA, which resulted in antiparallel to parallel (AP-P) switching, and then this switching was reversed (P-AP switching) when the current was swept back past -1 mA.

We have considered other potential mechanisms for this switching besides the SHE-ST. The Oersted field generated by the current can be ruled out because it has the polarity to *oppose* the switching that we observe, and it is small relative to the coercive field (0.7 mT at 1 mA, see SOM Section S3). We can also rule out the effect of any in-plane Rashba field (*32, 33*) that might be generated by $I_{Ta}$, because we measured the switching phase diagram of our three-terminal devices as the function of current and applied in-plane magnetic field. The result was as expected for thermally assisted anti-damping ST switching (*34*), and inconsistent with switching resulting from any type of current-generated effective field (SOM Section S4). In addition,

Suzuki *et al.* (*35*) have reported that any in-plane Rashba field, if it exists in a Ta/CoFeB/MgO multilayer, is oriented in the same direction as the Oersted field and would also act to oppose the switching that we observe. We conclude that the switching we measure is indeed the result of the ST exerted on the bottom MTJ electrode by the transverse spin current from the giant SHE in Ta.

By varying the current ramp rate (Fig. 3D) and using the standard model for thermally-activated ST switching (*34*), we determined both the zero-thermal-fluctuation ST critical currents and the energy barriers for the thermally activated AP-P and P-AP transitions. We found the two critical currents to be essentially the same, $|I_{c0}| = 2.0 \pm 0.1$ mA, and similarly for the energy barriers $U = 45.7 \pm 0.5\ k_B T$. The latter is not surprising but the former, while consistent with a SHE origin, is distinctly different from the case for ST switching by the spin polarized current generated by spin filtering within a spin valve or MTJ, where in general, $|I_{c0,P-AP}| \neq |I_{c0,AP-P}|$ due to, respectively, spin accumulation in the spin valve and the MTJ magnetoresistance behavior. The equivalence of the two critical currents for a SHE-ST switching device could be a significant technical advantage. From our measured values of $|I_{c0}|$ and using Eq. (1) with $\mu_0 M_{eff} = 0.76$ T (SOM Section S2), we determine $J_S/J_e$ for this device to be $0.12 \pm 0.04$ (SOM Section S1), in accord with our two other spin Hall angle measurements.

**Technology applications**. Improvements to this initial three-terminal SHE device can be very reasonably expected to result in significant reductions in the switching currents for thermally stable nanomagnets. Straightforward changes in the fabrication process that would reduce the width of the Ta microstrip close to the length of the long axis of the nanopillar would of course reduce $I_{c0}$ by a factor of 3 without affecting thermal stability. A further reduction in $I_{c0}$ could be achieved by reducing the demagnetization field of the FM free layer from 700 mT to $\leq$ 100 mT (*36, 37*). With such improvements $I_{c0}$ could be reduced to < 100 μA, at which point the three-terminal SHE devices would be competitive with the efficiency of conventional ST switching in optimized MTJs (*30, 31, 38*), while providing the added advantage of a separation between the low impedance switching (write) process and high impedance sensing (read) process. This separation solves the reliability challenges that presently limit applications based on conventional two-terminal MTJs while also giving improved output signals. Other three-terminal spin-torque devices based on conventional spin-filtering have been demonstrated previously (*39-42*), but the SHE-ST design can provide better spin-torque efficiency and is much easier to fabricate. Of course the discovery of materials with even larger values of the spin Hall angle than in *β*-Ta could also add to the competitiveness of SHE-ST.

In summary we have determined the strength of the SHE of *β*-Ta with three independent techniques and consistently find that the spin Hall angle is very large, $\theta_{SH} = 0.12 - 0.15$. The strength of the ST from the SHE remains consistent over FM layer thickness ranging from 1 to 4 nm, and is not sensitive to whether the FM layer is magnetized in-plane or out-of-plane. This demonstration of a giant SHE in Ta has important implications both for advancing the understanding of nanoscale magnetic phenomena involving Ta and other transition metal electrodes where the important role of the SHE has not always been appreciated [*e.g.*, (*43*)], and for enabling important technological applications. We show that the giant spin Hall effect in Ta can drive current-induced switching of either out-of-plane or in-plane-polarized magnetic samples, either of which might be used in magnetic memory or non-volatile logic applications. We have demonstrated an in-plane-polarized three-terminal SHE-ST device that is particularly promising for applications: compared to conventional MTJ structures it offers highly competitive spin torque efficiency together with isolation between the writing current and the reading current, allowing for a fully reliable write operation and a large signal read operation.

38. T. Kishi *et al.*, *In Proc. IEDM 2008, San Francisco, CA, 15–17 December 2008. New York, NY: IEEE. (doi:10.1109/IEDM.2008.4796680)*.
39. T. Kimura, Y. Otani, J. Hamrle, *Phys. Rev. Lett.* **96**, 037201 (2006).
40. T. Yang, T. Kimura, Y. Otani, *Nature Phys.* **4**, 851 (2008).
41. J. Z. Sun *et al.*, *Appl. Phys. Lett.* **95**, 109901 (2009).
42. P. M. Braganca *et al.*, *IEEE. Trans. Nanotech.* **8**, 190 (2009).
43. S. Fukami *et al.*, *Appl. Phys. Lett.* **98**, 082504 (2011).

**Acknowledgements**

We acknowledge support from ARO, DARPA, ONR, NSF/MRSEC (DMR-1120296) through the Cornell Center for Materials Research (CCMR), and NSF/NSEC through the Cornell Center for Nanoscale Systems. We also acknowledge NSF support through use of the Cornell Nanofabrication Facility/NNIN and the CCMR facilities.

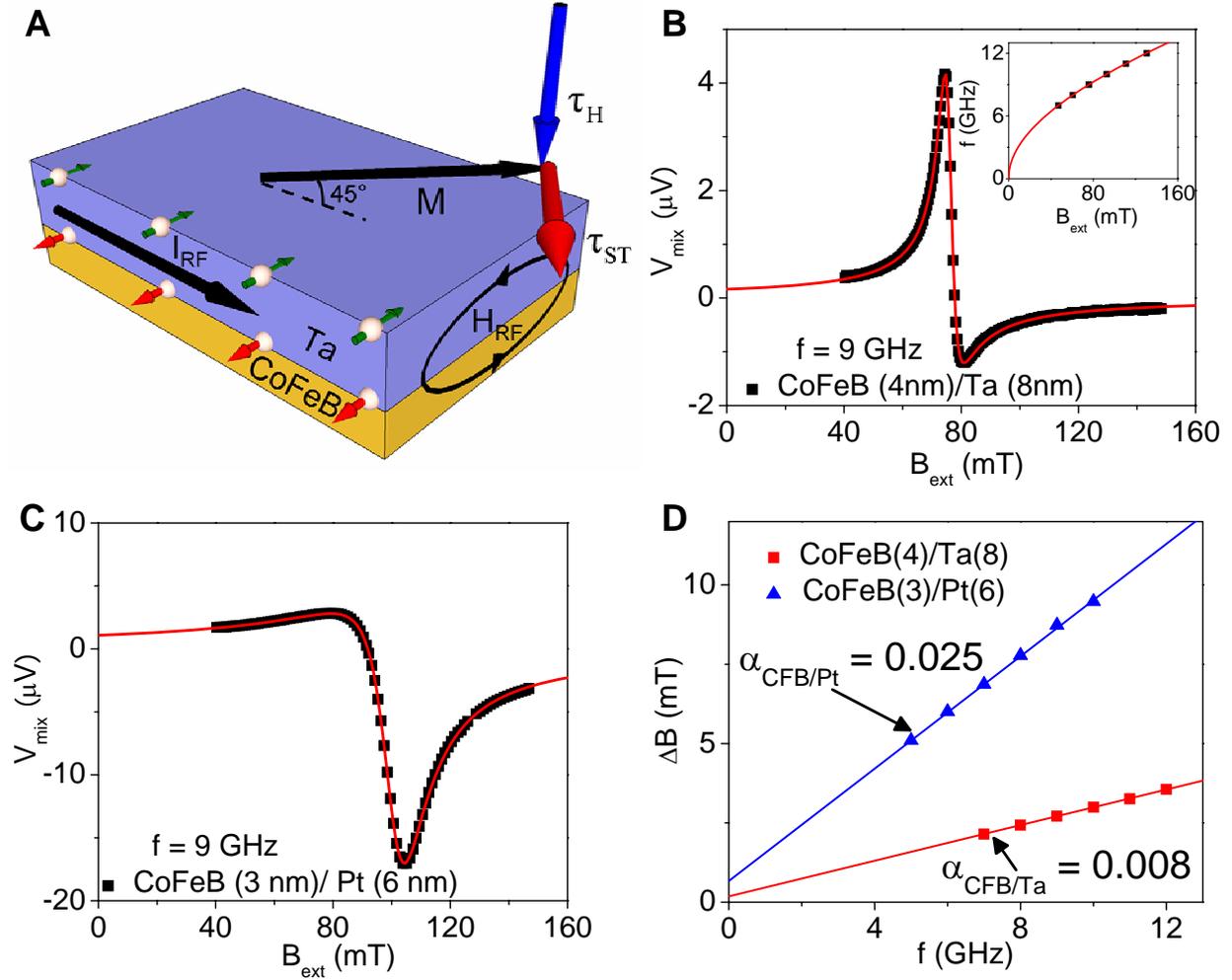

**Fig. 1.** ST-FMR induced by the spin Hall effect at room temperature. (**A**) Schematic of the sample geometry for the ST-FMR measurement. $I_{RF}$ and $H_{RF}$ represent the applied RF current and corresponding Oersted field. $\vec{\tau}_H = -\vec{M} \times \vec{H}_{RF}$ is the torque on the magnetization due to the Oersted field and $\vec{\tau}_{ST}$ is the spin transfer torque from the spin Hall effect. (**B**,**C**) Resonant lineshapes of the ST-FMR signals under a driving frequency $f = 9$ GHz for (**B**) CoFeB(4 nm)/Ta(8 nm) and (**C**) CoFeB(3 nm)/Pt(6 nm). The squares represent experimental data while the lines are fits to a sum of symmetric and antisymmetric Lorentzians. From the ratio of the symmetric and antisymmeteric peak components in (**C**), we determine the $J_S/J_e$ ratio for Pt to be

~0.07, consistent with earlier work (Ref. (*20*)). The inset of (**B**) shows how the frequency $f$ depends on the resonance magnetic field, in agreement with the Kittel formula (solid curve). (**D**) The resonance linewidth as determined from ST-FMR signals such as (**B**) and (**C**) at different resonance frequencies. The Gilbert damping coefficients $\alpha$ for Ta and Pt are calculated from the linear fits to the linewidth data.

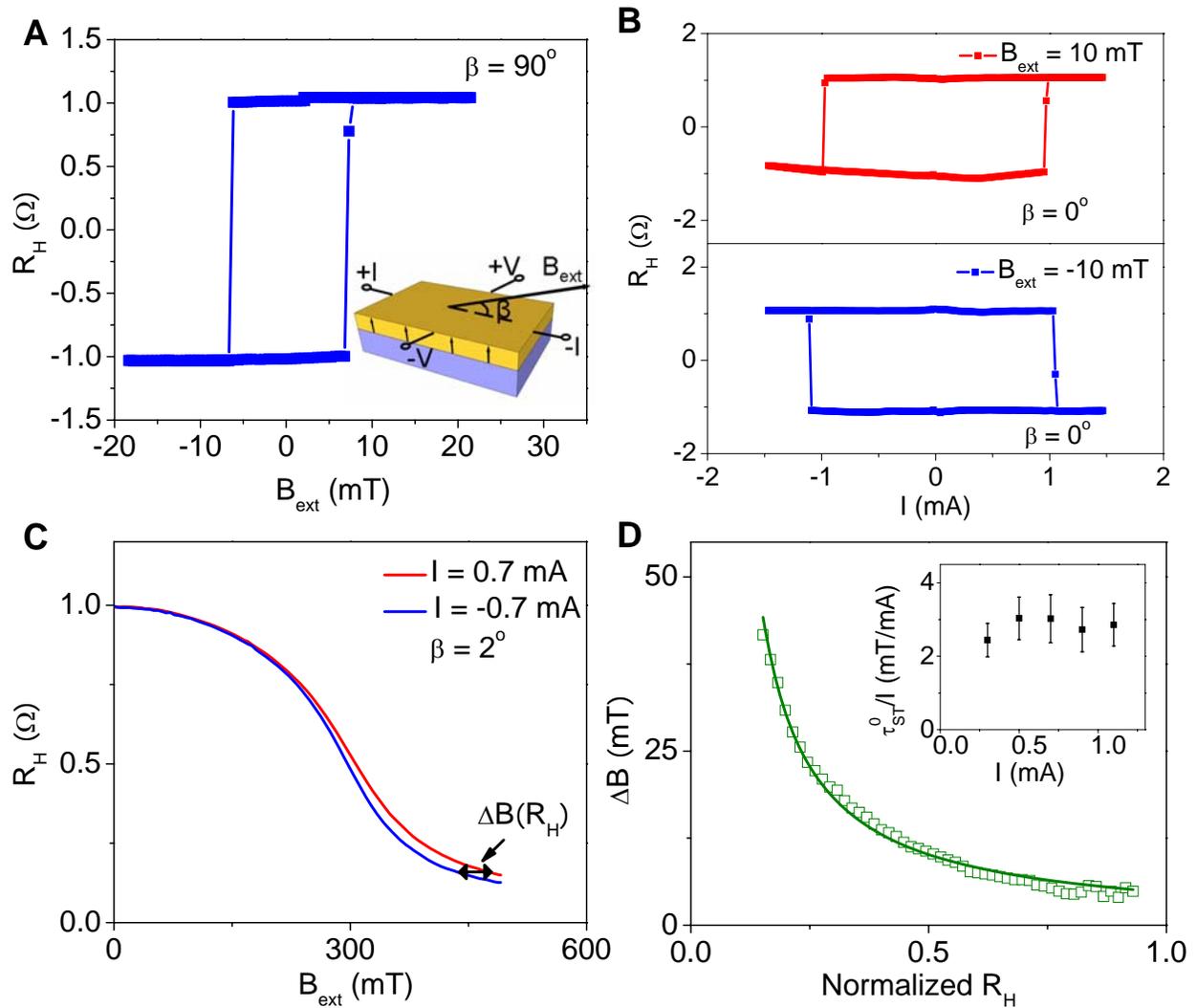

**Fig. 2.** Spin-Hall-effect-induced magnetic switching in a perpendicularly magnetized Ta/CoFeB/MgO/Ta film at room temperature. The sample was 2.5 μm wide and 3 μm long. (**A**) The anomalous Hall resistance $R_H$ as a function of magnetic field when $B_{ext}$ is applied along the easy axis (perpendicular to the film plane). Inset: schematic illustration of the device geometry used for the measurement. $B_{ext}$ is applied in the plane defined by the direction of current flow and the normal vector to the sample plane. $\beta$ is the angle between the direction of $B_{ext}$ and the applied

current. (**B**) Current-induced switching when $B_{ext}$ is parallel (top panel) or antiparallel (bottom panel) to the current direction defined as in (A) inset. In both panels, $\beta = 0°$. (**C**) $R_H$ vs $B_{ext}$ determined experimentally when the field is applied at the angle $\beta = 2°$. Constant currents of ±0.7 mA were applied to the sample while sweeping the field. (**D**) $\Delta B(R_H)$ as determined from the difference of the two data sets in (**C**). The line is a fit to the macrospin model. Inset: The values of $\tau_{ST}^0 / I$ determined at different bias currents.

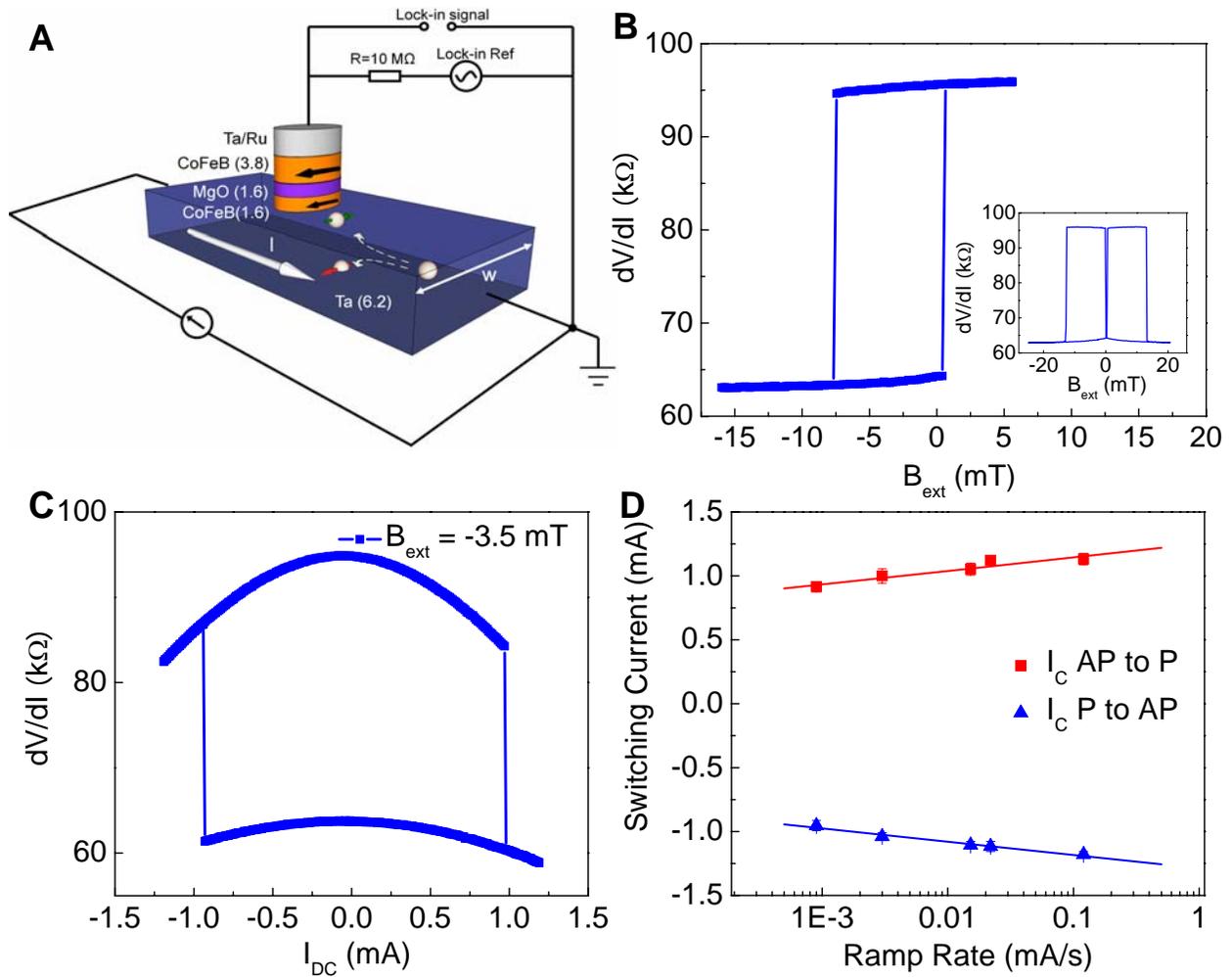

**Fig. 3.** Spin-Hall-effect-induced switching for an in-plane magnetized nanomagnet at room temperature. **(A)** Schematic of the three-terminal SHE device and the circuit for measurements. The direction of the spin Hall spin transfer torque is not the same as in Fig. 1A because the CoFeB layer now lies above the Ta rather than below. **(B)** TMR minor loop of the magnetic tunnel junction as a function of the external applied field $B_{ext}$ applied in-plane along the long axis of the sample. Inset: TMR major loop of the device. **(C)** TMR of the device as a function of applied DC current $I_{DC}$. An in-plane external field of -3.5 mT is applied to set the device at the center of the minor loop. **(D)** Switching currents as a function of the ramp rate for sweeping

current. The red squares indicate switching from AP to P and the blue triangles indicate switching from P to AP. Error bars are smaller than the symbol size.

**Supporting Online Materials**

# Spin torque switching with the giant spin Hall effect of tantalum

Luqiao Liu[*], Chi-Feng Pai[*], Y. Li, H. W. Tseng, D. C. Ralph and R. A. Buhrman

**Materials and Methods**

All of the films studied in our experiment were sputtered on thermally oxidized Si substrates in 2 mTorr Ar in a chamber with base pressure $< 2\times10^{-8}$ Torr. The microstrips used in the ST-FMR measurement were patterned using photolithography and Ar-ion milling. A brief plasma cleaning was employed before depositing Ta(10)/Pt(50) (thickness in nanometers) pads to make electrical contacts.

For the Hall bar geometries used in the perpendicular switching experiment, we defined the sample shapes using photolithography and ion milling, and after patterning deposited Ti(10)/Au(50) electrodes for the electrical contacts.

For the three-terminal devices, the 1μm-wide microstrip was defined by e-beam lithography and etched down to the substrate by Ar ion milling. A second aligned e-beam exposure was then used to define a rectangular 100 nm × 350 nm mill mask, and this pattern was transferred into the MTJ layer by ion milling. To ensure that the CoFeB free layer in the MTJ was fully patterned, we over-etched the Ta layer by ~ 2 nm. After the etching of the MTJ nanopillar, 70 nm of $SiO_2$ was deposited to provide electrical insulation between the bottom and the top electrodes. Finally, top and the bottom electrical contacts were defined by photolithography, and Ta(50)/Ru(50) was sputtered. The completed device was annealed at 280°C in vacuum for one hour before measuring to enhance the tunneling magnetoresistance. This annealing was not sufficient to diffuse the B out of the CoFeB and thus did not result in fully-crystallized CoFe.

**Supporting Text**

**Section S1. Calculation of critical current density $J_{C0}$ and the SHE efficiency $J_S/J_e$ from the zero-temperature switching currents of the three-terminal devices**

For a given measurement of the zero-temperature critical current $I_{c0}$, we can estimate the critical current density in the Ta layer required for spin torque switching as

$$J_{C0} = I_{c0} / (w_1 d_1 + w_2 d_2), \qquad (S1)$$

where $w_1$, $d_1$, $w_2$, and $d_2$ are defined as in Fig. S1. This equation assumes that the current flows uniformly in the CoFeB and Ta layers. This is a reasonable approximation since we determined the resistivities of the CoFeB and Ta from thicker thin-film samples in separate measurements and obtained values that are quite similar for the two materials: $\rho_{Ta} \approx 190$ µΩ-cm and $\rho_{CoFeB} \approx 170$ µΩ-cm. Using $d_1$ = 3.6 nm, $d_2$ = 4.2 nm, $w_1$ = 350 nm, $w_2$ = 1000 nm and $I_{c0}$ = 2 mA, we obtained $J_{C0} = 3.7 \times 10^7$ A/cm². We also measured the damping coefficient α of the 1.6 nm thick CoFeB film from Ta(6.2)/CoFeB(1.6)/MgO(1.6)/Ru(3.3) (thickness in nm) multilayers using the ST-FMR technique as is shown in the main text. From the linewidth of the resonance peaks at different frequencies, α is determined to be 0.021 ± 0.003, which is significantly larger than the damping coefficient of 4 nm thick CoFeB films that we studied in the main text. This number is consistent with previous work (*S1*), which showed that the damping coefficient for CoFeB has a strong thickness dependence when CoFeB is sandwiched between Ta and MgO and its thickness is reduced below 5 nm. With these values, we can estimate the $J_S/J_e$ ratio from Eq. (1) in the main text. Using $\alpha = 0.021$, $t = 1.6$ nm, $\mu_0 H_C = 4$ mT, $\mu_0 M_{eff} = 0.76$ T (see Section S2 below) and $M_S = 1.1 \times 10^6$ A/m (measured by magnetometry on large-area films), we find $J_S/J_e = 0.12 \pm 0.04$.

**Section S2. Determination of the demagnetization field of the 1.6 nm CoFeB layer**

Ta(6.2nm)/CoFeB(1.6nm)/MgO(1.6nm)/Ru(3.3nm) multilayers were sputtered onto thermally oxidized Si substrate under the same conditions as for the three-terminal device wafer and patterned into Hall bar microstrips using photolithography and ion milling. They were then vacuum annealed at 280°C for one hour. Two different size microstrips were formed, one with lateral dimensions 20 μm × 200 μm and the other 2.5 μm × 2.5 μm. Results of anomalous Hall resistance measurements are shown in Fig. S2, which indicate the demagnetization field ≈ 0.76 T at room temperature for both samples.

**Section S3. Determination of the direction of the Oersted field in the three-terminal devices**

We employed a Gauss meter to determine the direction of the external field ($B_{ext}$ in main text) and used Ampere's Law (right hand rule) to determine the direction of the Oersted field generated by current flow in the Ta microstrip. We determined that a positive (negative) current flowing through the Ta strip corresponded to an Oersted field that was aligned with the positive (negative) field direction as defined by $B_{ext}$ in Fig. 3B of the main text. Therefore the minor TMR loop shown in Fig. 3B of the main text shows that an Oersted field generated by a positive current in the Ta microstrip will tend to switch the free layer into the high resistance state, just as a sufficiently strong positive $B_{ext}$ does. This is opposite to what we observed in Fig. 3C of the main text, where a sufficiently strong positive current acts to switch the free layer into the low resistance states. Therefore we can conclude that the Orested field generated by the current flow in the Ta acts to oppose the switching driven by the spin torque arising from the SHE.

As an independent check on the direction of the Oersted field, we measured the TMR major loop under positive and negative currents flowing through the Ta microstrip. Since the fixed layer is only influenced by the Oersted field and not by the spin torque, we can determine the direction of the Oersted field from the current dependence of the fixed layer switching field in the major TMR loop. Since the fixed layer and the free layer are located on the same side of the microstrip, the Oersted field exerted on them should have the same direction. Fig. S3A and Fig. S3B show that the fixed layer switching fields shift negatively (positively) due to a positive (negative) current, which is in agreement with the results that we obtained above, *i.e.* the positive (negative) current generates an Oersted field in the positive (negative) direction. Note that the shifts of the free layer switching in Fig. S3 differ from those of the fixed layer because the SHE-ST shifts the switching fields of the free layer more strongly than does the Oersted field.

As a third and final check, we calculated the magnitude of the Oersted field generated by the current and compared it with the shift of the switching field of the fixed layer in Figs. S3A, B. By using the geometry described in Section S1, the Oersted field was calculated as (*S2*)

$$B_{Oersted} = \mu_0 J (d_1 + d_2)/2.  \tag{S2}$$

With $J = 1.83 \times 10^7$ A/cm$^2$ (corresponding to $I = 1$ mA), $d_1 = 3.6$ nm (CoFeB 1.6 nm/Ta 2 nm) and $d_2 = 4.2$ nm (Ta 4.2 nm) $B_{Oersted}$ was determined to be 0.9 mT, close to our measured result of $\approx 1.2$ mT (Fig. S3A and Fig. S3B). Similarly, the Oersted field exerted on the free layer can also be estimated considering that only the current in the Ta strip generates a net Oersted field on the free layer, since the Oersted field due to the current in the free layer itself averages out. The net Oersted field on the free layer is calculated to be ~ 0.7 mT.

**Section S4. Phase diagram of a three-terminal SHE-ST device**

Figure S4 shows the spin torque switching phase diagram as determined for a three-terminal MTJ device with lateral dimensions of 50 nm × 180 nm, formed on a 1 µm wide Ta microstrip. The smaller coercive field and lower switching currents of this MTJ device, in comparison to the one discussed in the main text, allowed a wider range of current to be applied without electrically damaging the device. The rhombohedral shape of the phase diagram (*i.e.*, with a shape that is closed on the top and bottom) is typical of that obtained from switching by a thermally-assisted spin torque mechanism (*S3*, *S4*), but cannot be explained by switching by an effective magnetic field. For a mechanism based on an in-plane effective field transverse to the current, the switching boundaries on this type of $B_{ext}$ vs. current graph would simply be two straight lines that do not meet on top and bottom.

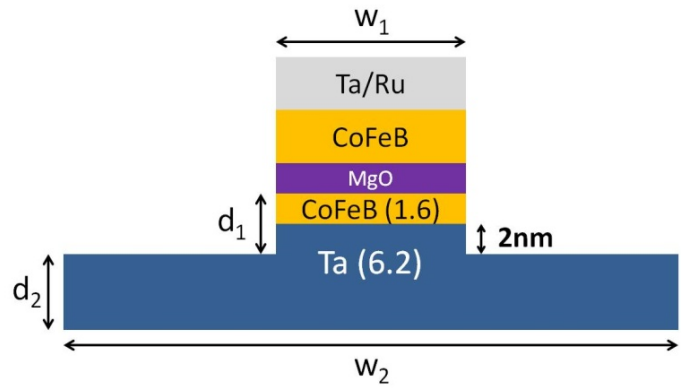

**Fig. S1.** Schematic cross section of the three-terminal device. The widths $w_1$ and $w_2$ are in the direction perpendicular to the current flow. $d_1$ is the thickness of CoFeB free layer plus the over-etched Ta thickness. $d_2$ is the thickness of the remaining Ta bottom layer. $w_1$ is the length of the long axis of MTJ pillar. $w_2$ is the width of Ta bottom layer.

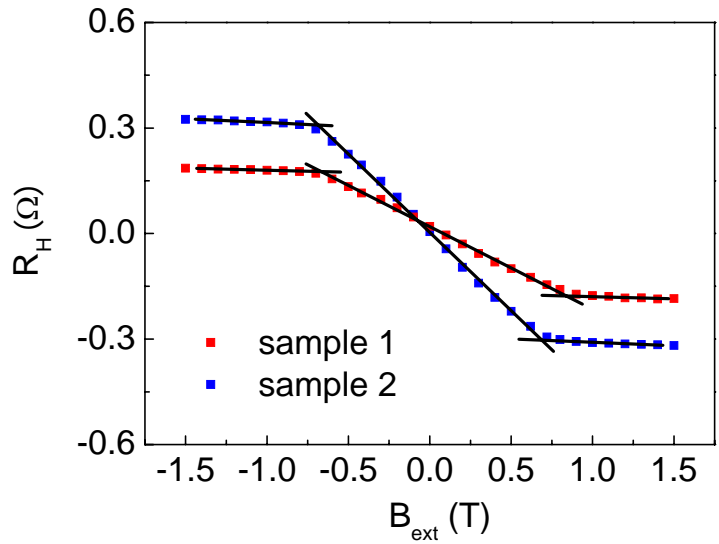

**Fig. S2.** Anomalous Hall resistance as the function of applied magnetic field for two different Ta(6.2 nm)/CoFeB(1.6 nm)/MgO(1.6 nm)/Ru(3.3 nm) samples. The magnetic field $B_{ext}$ is applied perpendicular to the film plane. Sample 1 and sample 2 have lateral dimensions 20 μm × 200 μm and 2.5 μm × 2.5 μm, respectively. The solid lines represent linear fits to the magnetization curves and the demagnetization field is given by the saturation field.

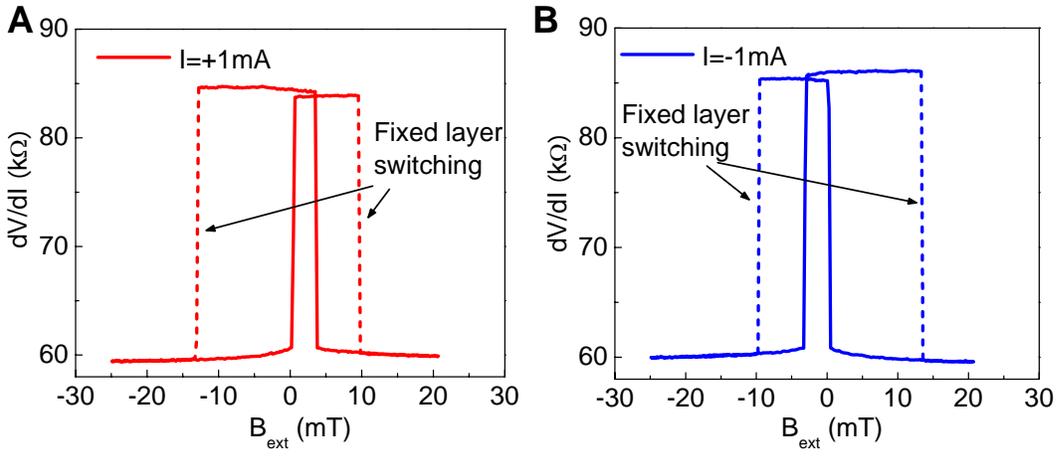

**Fig. S3.** TMR major loop of the three-terminal device with applied current (**A**) +1 mA and (**B**) -1 mA. The switching transitions for the fixed layer are labeled by dashed lines.

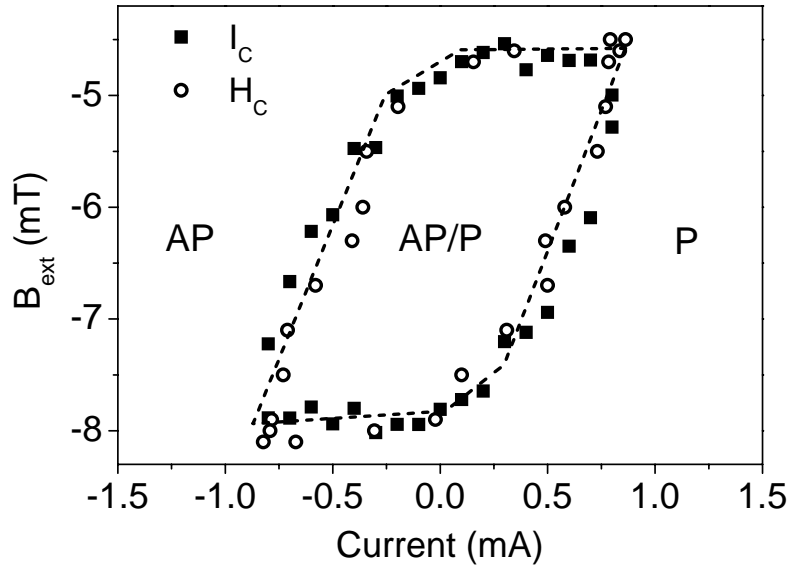

**Fig. S4.** Phase diagram of a three-terminal device, showing the boundaries for switching transitions between parallel (P), antiparallel (AP) and bistable (AP/P) states. The solid squares represent switching fields obtained from field scans at fixed current and the hollow circles represent switching currents obtained from current scans at fixed field. The dashed lines serve as a guide to eye. The dipole field from the fixed layer is ~ 6.3 mT. The rhombehedral shape of the bistable region is a signature of anti-damping switching by a spin transfer torque. In contrast, the boundaries for switching caused by a current-induced in-plane effective field would simply be straight lines on this type of $B_{ext}$ vs. current plot.